\documentclass[12pt]{article}
\usepackage{amsmath}
\usepackage{amsfonts} 
\usepackage{amssymb} 
\usepackage{hyperref}

\newenvironment{destaque}{\begin{quotation}\small\em}{\end{quotation}}

\date{}

\begin{document}

\title{\textbf{Maximally Localized States in Quantum Mechanics with a Modified Commutation Relation to All Orders}}
\author{Gl\'auber Carvalho Dorsch\thanks{Current address: \emph{Physics and Astronomy Department, University of Sussex, Brighton, BN1 9QH, UK}}\hspace{0.2cm}\thanks{G.Dorsch@sussex.ac.uk}  \hspace{0.3cm} and \hspace{0.3cm}Jos\'e Alexandre Nogueira\thanks{nogueira@cce.ufes.br}\\[0.1cm]\small\emph{Departamento de F\'isica, Centro de Ci\^encias Exatas}\\[-0.1cm]\small\emph{Universidade Federal do Esp\'irito Santo -- UFES}\\[-0.1cm]\small\emph{29075-910 -- Vit\'oria -- ES -- Brasil}}
\maketitle

\begin{abstract}
\begin{destaque}
 We construct the states of maximal localization taking into account a modification of the commutation relation between position and momentum operators to all orders of the minimum length parameter. To first order, the algebra we use reproduces the one proposed by Kempft, Mangano and Mann. It is emphasized that a minimal length acts as a natural regulator for the theory, thus eliminating the otherwise ever appearing infinities. So, we use our results to calculate the first correction to the Casimir Effect due to the minimal length. We also discuss some of the physical consequences of the existence of a minimal length, culminating in a proposal to reformulate the very concept of ``position measurement''.\\

\end{destaque}
\end{abstract}



\section{Introduction}

It is a remarkable fact that, despite the radical differences (not only formal, but in content as well) between the various approaches to quantum gravity, all of them seem to coincide in one prediction: the existence of a \emph{minimal length}, that is, a length scale below which the very concept of a length measurement loses its meaning. 

In quantum theory, the existence of a minimal length can be described as a nonzero minimal uncertainty $\Delta x_{min}$ in the measurement of position, which leads to a generalization of Heisenberg's uncertainty principle. Based on results obtained from string theory \cite{Amati:1988tn}, Kempf et al. \cite{Kempf:1994su} proposed a one-dimensional generalized uncertainty relation, which implements the appearance of a nonzero minimal uncertainty in position, of the form
\begin{equation}
	\label{string_uncertainty}
	\Delta x \Delta p \gtrsim \frac{\hbar}{2}  \left[ 1 +  \beta (\Delta p)^2 + \beta \langle \hat{P}  \rangle^{2}  \right] ,
\end{equation}
where $\beta$ is related to the minimal length. This generalized uncertainty relation corresponds to a modified commutation relation between the posittion operator and momentum operator of the form,
\begin{equation}
	\label{kempf}
	[\hat{X}, \hat{P}] = i\hbar(1+\beta \hat{P}^2)\ .
\end{equation}
This way a minimal length can be introduced into quantum mechanics by modifying its algebraic structure \cite{Maggiore1,Maggiore2}.

An important aspect of minimal length theories is that eigenstates of the position operator are no longer physical states, since for these states $\Delta x = 0 < \Delta x_{min}$. Consequently, we cannot work with the position representation $|x\rangle$ anymore, and, in order to recover some information on the spatial distribution of the system, we are forced to introduce the so-called ``quasiposition representation'', which consists in projecting the states onto the set of maximally localized states  $|\psi^{ml}_{\ \xi}\rangle$. The states $|\psi^{ml}_{\ \xi}\rangle$  are nothing else than states for which $\Delta x = \Delta x_{min}$, and are the generalization of the eigenstates $|x\rangle$ in the presence of a minimal length, in the sense that $|\psi^{ml}_{\ x}\rangle$ describes a system localized around $x$ with the best possible resolution.

An extremely interesting property of a minimal length is that its existence limits explicitly the resolution of small distances in the spacetime, providing a natural regulator of Quantum Field Theory.

In this work, we do not discuss the existence or the probable origins of a minimal length. Instead, we assume it as existent, and we introduce it \emph{ad hoc} in the formalism of Quantum Mechanics in order to investigate its consequences, which concepts lose their meaning in its presence, and which new concepts eventually emerge. We do this, as mentioned above, by modifying the uncertainty relation in order to introduce a nonzero minimal value for $\Delta x$, which, being a limitation for the localizability of the physical systems, acts as a minimal length. However we emphasize that (\ref{string_uncertainty}) is expressed only to first order of the minimum length parameter, and nothing prevents the existence of higher order terms. Our proposal is to extend (\ref{kempf}) to all orders of the minimum length parameter $\beta$, and to construct the maximally localized states associated to it. As an application we propose to calculate the Casimir force expanding the field operator in those maximally localized states.

The article is organized in the following way. In section \ref{comp_min_MQ} the proposed modified Heisenberg algebra is presented and we use Jensen's inequality to show that it indeed leads us to a $\Delta x_{min}>0$, as it should to fulfil our goal. We then build the corresponding states of maximal localization in section \ref{Hilbert_space}. The conceptual issues that arise due to the absence of the position representation in this framework are tackled in section \ref{Discussions}. Another consequence of a minimal length worth being mentioned is that, because it extinguishes the notion of local interactions, it acts as a natural regulator parameter in QFT. In section \ref{regulator} we calculate the 1-1 dimensional Casimir Effect, emphasizing how the minimal length naturally regularizes it. We present our conclusions and outlooks in section \ref{conclusions}.


\section{Minimal Length in Quantum Mechanics}
\label{comp_min_MQ}

One way to introduce a minimal length in Quantum Mechanics is to modify the uncertainty relations so that a nonzero minimum value for $\Delta x$ emerges, which, as a limitation on the localizability of a system, plays this desired role. Since

\begin{equation}
	\label{eqIncertezaGeneralizada}
	\Delta x \Delta p \geq \frac{|\langle [\hat{X},\hat{P}] \rangle |}{2},
\end{equation}
a modification of the uncertainty relations requires a modification in the algebra of the operators, and this implies in a complete reformulation of the structure of the theory.

In this work, we try to follow the approach of Kempf, Mangano, and Mann (KMM)\cite{Kempf:1994su}, but with an algebra that takes into account corrections of all orders in the minimal length parameter, namely
\begin{equation}
	\label{algebraproposta}
	[\hat{X},\hat{P}]=i\hbar e^{\beta \hat{P}^2}\ .
\end{equation}
Note that, to first order of $\beta$ (the parameter related to the minimal length) we recover (\ref{kempf}).

The first thing we ought to do is to show that we indeed obtain a minimum value for $\Delta x$, which is our goal in the first place. From (\ref{eqIncertezaGeneralizada}) and (\ref{algebraproposta}) we have

\begin{equation}
	\label{incertezamodificada}
	\Delta x \Delta p \geq \frac{\hbar}{2}\langle e^{\beta \hat{P}^2}\rangle\ .
\end{equation}
Now, the so-called Jensen's Inequality\cite{Jensen} guarantees that, if $\phi$ is a real convex function, then $\langle \phi(\hat{X}) \rangle \geq \phi(\langle \hat{X}\rangle)$. Since $e^{x^2}$ satisfies these requirements it follows that
\begin{equation}
	\Delta x \Delta p \geq \frac{\hbar}{2}e^{\beta \langle \hat{P}^2 \rangle} = \frac{\hbar}{2}e^{\beta {\langle \hat{P} \rangle}^2}e^{\beta (\Delta p)^2}\ ,
\end{equation}
implying
\begin{equation}
	\label{deltaxminimo}
	\Delta x \geq \hbar \sqrt{\beta}\sqrt{\frac{e}{2}}e^{\beta {\langle \hat{P} \rangle}^2} \geq \hbar\sqrt{\beta}\sqrt{\frac{e}{2}}\ ,
\end{equation}
that is, (\ref{algebraproposta}) implies the existence of a $\Delta x_{min} \geq \hbar\sqrt{\beta}\sqrt{\frac{e}{2}}$\footnote{In B. Bagchi and A. Fring\cite{Bagchi:2009wb}, this result was shown using a numerical method.}.


\section{Hilbert Space}
\label{Hilbert_space}

The next step in this (re)development of the formalism is to build the Hilbert Space $\mathcal{H}$ to which the physical states belong.

An important aspect of minimal length theories is that eigenstates of the position operator are no longer physical states, since for these states $\Delta x = 0 < \Delta x_{min}$. Consequently, we cannot work with the position representation $|x\rangle$ anymore. In ``ordinary Quantum Mechanics''\footnote{We use this term to refer to Quantum Mechanics without a minimal length.} we also had $|x\rangle \notin \mathcal{H}$, for these vectors have infinite norm\footnote{The theory is already ``warning us'' about the problems related to the concept of perfect localizability.}, but in this case we could work around the problem by defining $|x\rangle$ as a limit of the sequence of states $|\xi_{\ x}^{\Delta x}\rangle$ --- for which $\langle\hat{X}\rangle=x$ and whose uncertainty is $\Delta x$ --- when $\Delta x \rightarrow 0$\footnote{Rigorously speaking, this limit is also ill defined, since it suffers from the same problems the vectors $|x\rangle$ do. However, we can use this procedure of taking a limit to define $\langle x | \hat{A} | \psi \rangle \equiv \lim_{\Delta x \rightarrow 0}\langle \xi_{\ x}^{\Delta x} | \hat{A}|\psi \rangle$ for any operator $\hat{A}$ and any vector $|\psi\rangle$. That is, we first evaluate $\langle \xi_{\ x}^{\Delta x} | \hat{A}|\psi \rangle$ and, then, take the limit, which is well-defined for an adequate choice of the sequence $|\xi_{\ x}^{\Delta x}\rangle$.}. In our present case, not even that is possible, since we cannot take the limit $\Delta x \rightarrow 0$.

In the absence of the position representation, the simplest alternative is to work with the momentum representation $|p\rangle$\footnote{When both representations are absent, due to a minimal uncertainty also in the momentum, an alternative is to work with the Bargmann-Fock representation. A detailed study of such case can be seen in A. Kempf\cite{Kempf:1996ss}.}. The $\hat{X}$ and $\hat{P}$ operators in this representation are given by
\begin{equation}
	\begin{split}
		\label{novosoperadores}
		\langle p | \hat{P} | \psi \rangle &= p \psi(p), \\
		\\	
		\langle p | \hat{X} | \psi \rangle &= i\hbar e^{\beta p^2}\frac{\partial}{\partial p}\psi(p)\ ,
	\end{split}
\end{equation}
where the exponential factor in the definition of $\hat{X}$ is required so that (\ref{algebraproposta}) be satisfied. Note that this implies a modification of the Schr\"odinger equation, which means the introduction of the parameter $\beta$ affects the evolution of every quantum system.

These operators are hermitean under the inner product
\begin{equation}
	\label{inner_product}
	\langle \phi | \psi \rangle = \int_{-\infty}^{\infty}{e^{-\beta p^2}\phi^*(p) \psi(p) dp}\ ,
\end{equation}
and from this we see that the measure of momentum space is modified, i.e., 
\[	dp \rightarrow e^{-\beta p^2}dp\ .\]
This modification is not unexpected. The existence of a minimal length implies in an upper limit to the wave-number $k$, as we have already seen, so that any integral in $k$-space has finite integration limits. The same does not happen in momentum space, which is unlimited. Therefore, in a change of integration variables from $k$ to $p$, the Jacobian must be a decreasing function that compensates this difference in the integration limits. It is this new damping exponential factor in the measure that makes Quantum Field Theories naturally finite in the presence of a minimal length.

Finding another representation to the Hilbert Space is far from being the only difficulty to be overcome. Since the eigenvectors $|x\rangle$ of $\hat{X}$ do not belong to $\mathcal{H}$ anymore, this operator ceases to be an observable, even though it is still hermitean\footnote{An hermitean operator is an observable if its eigenvectors form a basis of the Hilbert space.}. This immediately raises questions like: what is the meaning of the expression ``position measurement'' in this formalism (if it still has any meaning, since the operator associated to any measurable quantity must be an observable)? And how could we recover information on the propagation of the system in space? For example, how could we calculate the probability of finding the system in a certain spatial region? 

These questions cannot simply be ignored and left unanswered, for three main reasons. First, because the formalism we now develop must coincide with the results of ordinary QM in the limit $\beta \rightarrow 0$, and, therefore, there must be some notion of spatial localization of the quantum system in this modified QM that, in this limit, gives the definition with which we are familiar. Secondly, the value of $\Delta x$ associated to each state plays a central role in our work, since it is through a $\Delta x_{min}$ that we introduce a minimal length in the theory; thus, it is necessary to guarantee that such $\Delta x$ exists, that it makes sense even when $\hat{X}$ ceases to be an observable, and to interpret its physical meaning in this case. Finally, because to understand some of the conceptual changes imposed by the existence of a minimal length is precisely one of the goals of this work.


\subsection{Maximally Localized States}

To answer these questions, we will use the concept of ``maximal localization states'' introduced in A. Kempf, G. Mangano and R. B. Mann\cite{Kempf:1994su}, which are nothing else than states for which $\Delta x = \Delta x_{min}$, i.e., states $|\psi^{ml}_{\ \xi}\rangle$ obeying

\begin{equation}
	\begin{split}
 		\langle \psi^{ml}_{\ \xi}|(\Delta \hat{X})^2|\psi^{ml}_{\ \xi}\rangle &= (\Delta x_{min})^2 \\
 		\langle\psi^{ml}_{\ \xi}|\hat{X}|\psi^{ml}_{\ \xi}\rangle &= \xi\ .
 	\end{split}
\end{equation}

The state $|\psi^{ml}_{\ x}\rangle$ is the generalization of the eigenstate $|x\rangle$ in the presence of a minimal length, in the sense that $|\psi^{ml}_{\ x}\rangle$ describes a system localized around $x$ with the best possible resolution (which is not infinite in the present case). Indeed, $|\psi^{ml}_{\ x}\rangle \rightarrow |x\rangle$ when $\Delta x_{min}\rightarrow 0$, as we shall see. It is this property that makes them useful in trying to answer the above raised problems.

In what follows, we shall try to find an expression for these states in momentum representation.


\subsubsection{Na\"ive Approach}

The first attempt, which we call the ``na\"ive approach'' for reasons that will become apparent in a moment, would be to repeat the Kempf, Mangano, and Mann (KMM) prescription\cite{Kempf:1994su}. In their case, the starting point is the commutation relation given by Eq. (\ref{kempf}), which leads to the uncertainty relation \[ \Delta x \geq \frac{\hbar}{2\Delta p}(1+\beta\langle P \rangle^2)+\frac{\hbar \beta}{2}\Delta p . \] 
It is then clear that $\Delta x$ will be a minimum when the above relation is an equality, which occurs only if
\begin{equation}
	\label{condicaoincertezaigualdade}
	\left[\hat{X} - \langle \hat{X} \rangle + \frac{\langle [\hat{X},\hat{P}] \rangle}{2\langle (\Delta\hat{P})^2\rangle}(\hat{P} - \langle \hat{P} \rangle) \right]|\psi_{squeezed} \rangle = 0.
\end{equation}
We call states that obey this condition \emph{squeezed states}, $\mathcal{H}_{squeezed}$ being their corresponding subspace. The squeezed state with the minimal value for $\Delta x$ is the maximally localized stated we are looking for.

We could then follow the same reasoning and, therefore, apply the same equation above to try to find our maximally localized states. Putting (\ref{algebraproposta}) into (\ref{condicaoincertezaigualdade}), projecting on $\langle p|$ and solving the differential equation we find
\begin{equation}
	\label{gaussianageneralizada}
	\psi_{squeezed}(p) \propto e^{-ik \langle \hat{X} \rangle} exp \left[\frac{\Delta x}{\hbar \Delta p}\left(\frac{e^{-\beta p^2} - 1}{2\beta} + \hbar k \langle \hat{P} \rangle \right)\right]\ ,
\end{equation}
where
\begin{equation}
   \label{relationkp}
	\hbar k(p) = \int_{0}^{p}{e^{-\beta q^2}dq}\ .
\end{equation}
Recalling that the Gaussian state is exactly the state for which the uncertainty relation is an equality, we see that $\psi_{squeezed}(p)$ is the generalized Gaussian wave function in the presence of a minimal length. Indeed, it is easy to see that we recover the familiar Gaussian function in the limit $\beta \rightarrow 0$. Moreover, the oscillatory factor $e^{-ik \langle \hat{X} \rangle}$ in Eq. (\ref{gaussianageneralizada}) has the form of a plane wave, which allows us to interpret $k(p)$ as a wave-vector. We point out that the functional relation (\ref{relationkp}) between the wave-vector $k$ and the momentum $p$ is not new. It has been used by S. Hossenfelder in order to introduce a parametrization of the  minimal length effects\cite{Hossenfelder:2004}.

We can use this expression for $k(p)$ to estimate a relation between the parameter $\beta$ and the minimal length $\Delta x_{min}$. To begin with, we note that, when $p\rightarrow\infty$, the wavelength $\lambda$ of the wave associated to the particle approaches a minimum $\lambda_{min}$, which must be greater than or equal to $\Delta x_{min}$. Hence,
\[	\frac{2\hbar\pi}{\Delta x_{min}}\geq\frac{2\hbar\pi}{\lambda_{min}} = \lim_{p \to \infty} \hbar k(p) \rightarrow \int_{0}^{\infty}{e^{-\beta q^2}dq} = \frac{\sqrt{\pi}}{2\sqrt{\beta}}, \]
which implies
\[	\Delta x_{min} \leq 4\hbar \sqrt{\pi \beta}\ .\]
From this and (\ref{deltaxminimo}) it follows that
\begin{equation}
 \sqrt{\frac{e}{2}} \leq \frac{\Delta x_{min}}{\hbar \sqrt{\beta}} \leq 4\sqrt{\pi}\ , 
\end{equation}
i.e., $\Delta x_{min} = a\hbar \sqrt{\beta}$,
with
\begin{equation}
	\label{numerical_estimative}
	 1.17\approx \sqrt{\frac{e}{2}}\leq a \leq 4\sqrt{\pi}\approx 7.09.
\end{equation}

From (\ref{gaussianageneralizada}), a maximally localized state would be obtained by imposing that $\Delta x=\Delta x_{min}$ (which implies $\langle \hat{P} \rangle=0$, see Eq. (\ref{deltaxminimo})). Thus,
\begin{equation}
	\label{ondaplanageneralizada}
	\psi^{ml}_{\ \xi}(p) \propto e^{-ik \xi} exp \left(\frac{\Delta x_{min}}{\hbar \Delta p^{min}}\frac{e^{-\beta p^2} - 1}{2\beta}\right)\hspace{1cm}\emph{(na\"ive result)}
\end{equation}
where $\Delta p^{min}$ is the value of $\Delta p$ that minimizes $\Delta x$\footnote{$\Delta p^{\min}$ has nothing to do with a minimal uncertainty in the momentum. The notation is admittedly bad, but we cannot think of a better one.}. Equation (\ref{ondaplanageneralizada}) would, then, be the maximally localized state we are looking for. 

However, as it was pointed out by S. Detournay, C. Gabriel and P. Spindel\cite{Detournay:2002fq}, the KMM prescrition is not appropriate to construct maximal localization states for commutation relations like Eq. (\ref{algebraproposta}). So, the result (\ref{ondaplanageneralizada}) is not correct. The flaw lies in the premise that we must look for maximally localized states among squeezed states, as was done in A. Kempf, G. Mangano and R. B. Mann\cite{Kempf:1994su}. While this prescription works well with their assumed commutation relation, in our case the right-hand side of the uncertainty relation is a complicated expression whose explicit form we do not know, and whose value depends on the state on which it is evaluated. It may well be possible to find a state outside $\mathcal{H}_{squeezed}$ (i.e., a state for which Eq. (\ref{incertezamodificada}) is satisfied as an \emph{in}equality), but whose $\Delta x$ is still smaller than inf$\left\{\langle \psi |\sqrt{(\Delta \hat{X})^2}|\psi\rangle,\ |\psi\rangle \in \mathcal{H}_{squeezed}\right\}$.


\subsubsection{The DGS Approach}

Having detected the problem of the KMM prescription when applied to general commutation relations, Detournay, Gabriel and Spindel (DGS)\cite{Detournay:2002fq} go on to define an alternative and more abrangent one, which is suitable to our particular case and which we now briefly describe.

Our goal is to find the states $|\psi^{ml}_{\ \xi}\rangle$ belonging to a certain subspace $\mathcal{H}_{phys}$\footnote{Note that $\mathcal{H}_{phys}$ is larger than $\mathcal{H}_{squeezed}$.} of the physical states and satisfying
\begin{equation}
	(\Delta x_{min})^2=\text{min}\{\langle \psi^{ml}_{\ \xi} | \hat{X}^2 - \xi ^2|\psi^{ml}_{\ \xi}\rangle\}\equiv \mu^2 ,
\end{equation}
where 
\begin{equation}	
	\label{vm}
	\xi = \langle \psi^{ml}_{\ \xi} | \hat{X} | \psi^{ml}_{\ \xi}\rangle .
\end{equation}
In momentum representation, this corresponds to 
\begin{equation}
	\left[-\left(\hbar e^{\beta p^2}\partial_p\right)^2 - \xi^2 + 2a\left(i\hbar e^{\beta p^2}\partial_p - \xi\right) -\mu^2\right]\psi^{ml}_{\ \xi}(p)=0,
\end{equation}
where $a$ is a Lagrange multiplier introduced to account for Eq. (\ref{vm}). Imposing certain conditions to guarantee that the state indeed belongs to $\mathcal{H}_{phys}$\footnote{We add, for example, the condition that $\langle \psi^{ml}_{\ \xi} | \hat{V}(p) | \psi^{ml}_{\ \xi}\rangle$ is finite for some unbounded observable $\hat{V}$, e.g. the energy. In such case, all we are saying is that the energy of a physical state must be finite.} we arrive at the solution
\begin{equation}
	\label{DGSstate}
	\psi^{ml}_{\ \xi}(p)\propto e^{-i\xi k(p)}sin\left[n\Delta x_{min}\left(k(p)+\frac{1}{2\hbar}\sqrt{\frac{\pi}{\beta}}\right)\right],
\end{equation}
with $n\in \mathbb{N}$ and 
\begin{equation}
	\label{minimal_length}
	\Delta x_{min}=\hbar\sqrt{\beta \pi}.
\end{equation}

This gives us an \emph{exact} relation between the parameter $\beta$ and the minimal length. Notice that $\frac{\Delta x_{min}}{\hbar\sqrt{\beta}}\approx 1.77$, in agreement with (\ref{deltaxminimo}) and (\ref{numerical_estimative}). Notice, moreover, that we have again an exponential with the form of a plane wave, with $k(p)$ playing the role of the wave-number, as it should be. 

Finally, it is easy to see that $\displaystyle{lim}_{\beta\rightarrow 0}\psi^{ml}_{\ \xi}(p)\propto e^{-i \xi \frac{p}{\hbar}}$, which proves that $|\psi^{ml}_{\ \xi}\rangle$ is indeed a generalization of the eigenvectors $|x\rangle$, as was claimed in the beginning of this section.



\section{Conceptual issues}
\label{Discussions}

We have seen that the existence of a minimal length raises many conceptual problems related to how to recover information on the localization of the system in space. They emerge because, in ordinary QM, all these informations are obtained by use of the eigenvectors $|x\rangle$ of $\hat{X}$, and these vectors do not belong to the Hilbert Space of the theory when a minimal length is present.

However, we have shown above that the maximally localized states are a generalization of these $|x\rangle$ to the formalism we now develop, and this might enable us to try to use them to answer the conceptual questions posed above. This is what we will do now.


\subsection{Measurements in Quantum Mechanics}

The first, and most fundamental of these questions, refers to the meaning of the expression ``position measurement'' in the presence of a minimal length. Let us investigate further what is really meant by this expression to understand where the problem is, and to try to solve it.

Let $\hat{A}$ be the observable associated to some physical quantity $\mathcal{A}$, $\sigma_A$ be its spectrum, $\mathcal{H}_a$ be the eigenspace related to the eigenvalue $a\in \sigma_A$, and $\hat{P}_a$ be the orthogonal projector onto $\mathcal{H}_a$. 

In Quantum Mechanics, we are familiar with the notion that a measurement of $\mathcal{A}$ is a perturbation acting on the system that causes it to collapse to one of the eigenspaces $\mathcal{H}_a$, whose corresponding eigenvalue is, then, the value obtained in the measurement. To the measurement process it is associated the projector $\hat{P}_{a}=\sum_{i=1}^{k_a}|a^i\rangle\langle a^i|$, $k_a$ being the degeneracy of the obtained eigenvalue.

But there is a subtlety involved in this definition: it is valid only in the restricted case of a \emph{sufficiently selective measurement}, i.e. a measurement whose result is a \emph{single} value $a$. In general, however, a measurement causes the system to collapse not to a single eigenspace, but to a certain union of them, and the projector associated to the process is generally written as $\hat{P}=\sum_{a\in\sigma_A}c_a \hat{P}_a$, where $c_a$ is related to the ``intensity'' with which each eigenvalue is obtained in the measurement.

All this becomes clearer (and more evident) when $\hat{A}$ has a continuous spectrum. In this case, it is impossible to obtain a single value in a measurement, for it would require a device with an \emph{infinite resolution}, and such thing does not exist\footnote{This is indicated by the very formalism of the theory. If there were such a device, the state of the system afterwards would be a vector $|a\rangle$ that does not belong to $\mathcal{H}$, as we already pointed out for the particular case of position eigenvectors $|x\rangle$.}. In such case, every measurement results in an interval $I^{\delta a}_{a}$ centered in $a$ and of length $\delta a$, and we then loosely say that $a$ is the value measured with a resolution $(\delta a)^{-1}$. The associated projector is given by 
\begin{equation}
	\label{projetor_continuo}
	\int_{-\infty}^{\infty}c(a')\hat{P}_{a'}da',
\end{equation}
where $c(a')$ is a function whose exact expression depends on the measuring apparatus, but which, in general, has a maximum at $a$ and is approximately zero when $|a'-a|\gtrapprox \delta a/2$. In the case of a position measurement, say, performed with a photographic plate that becomes brighter in the region with which the incident particle interacts, the $c(x')$ gives the intensity of this brightness as a function of the coordinate $x'$ of the plate. Since we can hardly tell the exact form of this function, we formulate the working hypothesis that the apparatus is a \emph{perfect filtering device}, which means the collapsed wave-function is exactly zero outside $I^{\delta a}_{a}$, and remains unaltered otherwise. That is, 
\begin{equation}
	\label{perfect_filter}
	\displaystyle{c(a') = \left\{ \begin{array}{ll} 0,\ &a' \notin I^{\delta a}_{a} \\
							 1,\ &a' \in I^{\delta a}_{a} 
					\end{array} \right.}
\end{equation}
Under these conditions, (\ref{projetor_continuo}) assumes the form 
\begin{equation}
	\label{projector_perfect_filter}
	\int_{a-\frac{\delta a}{2}}^{a+\frac{\delta a}{2}}\hat{P}_{a'}da'
\end{equation}
which is exactly the form of a projector associated to a measure of a degenerate eigenvalue, whose eigenvectors are $\{|a'\rangle, a'\in I^{\delta a}_{a}\}$. This allows us to reinterpret an \emph{insufficiently selective measurement} in the following way. 

Let the interval $I^{\delta a}_{a}$ be given in advance. Define an operator $\mathcal{O}_a^{\delta a}$ whose eigenvalues are $0$ and $1$, and whose eigenvectors associated with the eigenvalue $1$ are the vectors for which a measurement of $\hat{A}$ yields values contained in $I^{\delta a}_{a}$, while the eigenvectors associated to $0$ are the vectors for which a measurement of $\hat{A}$ yields values that are not in such interval. Since $\hat{A}$ and $\mathcal{O}_a^{\delta a}$ share the same eigenvectors --- only the associated eigenvalues change ---, $\hat{A}$ being an observable implies that $\mathcal{O}_a^{\delta a}$ is an observable, and, therefore, there is a physical quantity associated to it. Now, note that the projector onto the eigenspace $\mathcal{H}_1$ of this operator is precisely (\ref{projector_perfect_filter}), which means that to measure $\hat{A}$ and obtain $I^{\delta a}_{a}$ is equivalent to measure $\mathcal{O}_a^{\delta a}$ and obtain $1$.

This means that we can interpret a position measurement as a question of ``whether or not the system is localized in a certain predefined region of space''. A position measurement apparatus with resolution $(\delta x)^{-1}$ is, thus, nothing more than a set of detectors with length $\delta x$, each of which changes its state (say, emits a beep) when it interacts with the particle. 


\subsection{Position Measurement with a Minimal Length}

In the presence of a minimal length, the eigenvalues of the position operator do not belong to the Hilbert Space anymore, and, as a consequence, $\hat{X}$ loses its status of an observable. Since there are no more eigenspaces $\mathcal{H}_x$, nor projectors $P_x$ onto such spaces, it is simply impossible to speak of a position measurement --- even of a measurement with $\Delta x>\Delta x_{min}$, which would not violate our initial postulate. 

Nevertheless, we should still find some way to work around this problem and recover information on the spatial configuration of the system, otherwise the theory would be incomplete, even incapable of describing many systems in which the essential informations are on spatial localization, like a particle moving in a cloud chamber, for example. 

This can indeed be done using the alternative interpretation of a position measurement given above. We say that to ``measure'' the position of a system with the greatest possible resolution is to make it interact with a properly regulated apparatus (an array of detectors each with length $\Delta x_{min}$) that indicates if this interaction occurred in a previously determined spatial region, centered in $\xi$ and with length $\Delta x_{min}$, or not\footnote{For a similar interpretation, generalized to the covariant case (in which we measure the localization of the system not only in Space, but in Spacetime) see M. Reisenberger and C. Rovelli\cite{Reisenberger:2001pk}.}. In case affirmative, the perturbation causes the system to collapse to the respective Maximal Localization State, $|\psi^{ml}_{\xi}\rangle$, and the ``projector'' associated to the ``measurement'' is 
\begin{equation}
	\label{projetorML}
	\hat{P}^{ml}_{\xi} = |\psi^{ml}_{\xi}\rangle\langle\psi^{ml}_{\xi}|.
\end{equation}

Strictly speaking, we cannot say this is a measurement\footnote{Hence the quotes in the above paragraphs.}, since we have not presented an observable associated with it. But it definitely is a way to recover information on the spatial localization of a system, and this was, after all, all we were looking for.

Under one aspect, this definition is even less problematic than the one of ordinary QM. In that case, the correct form of the projector associated to the measure is (\ref{projetor_continuo}), and the coefficients $c(x')$ are unknown. We generally solve this difficulty by making the assumption that leads to (\ref{projector_perfect_filter}), but this assumption is physically incorrect, not only because there are no perfect filters in reality, but, much graver than that, because the action of this projector on a state may result in a discontinuous wave-function, which is absurd. In the present case no such problems arise: the ``projector'' is known to be (\ref{projetorML}), and the resulting state after the ``measurement'' is simply $|\psi^{ml}_{\xi}\rangle$.


\subsection{Probability and Mean Values}

The Maximal Localization States can also be used to calculate the probability for a certain system, whose state is described by $|\phi\rangle$, to be found around the point $\xi$ when its position is ``measured'' with the greatest possible resolution. This probability is, of course, given by
\[ \mathcal{P}^{ml}_{\ \xi} = |\langle\psi^{ml}_{\xi}| \phi \rangle|^2. \]
The function
\[ \phi(\xi) \equiv \langle\psi^{ml}_{\xi}| \phi \rangle \]
is the generalization of the wave-function of ordinary Quantum Mechanics, and gives us an useful expression for the spatial distribution of the system. 

There is yet another way to recover information on how the system is localized in space, which is by calculating the mean values related to the operator $\hat{X}$, like 
\begin{equation}
	\label{valoresmedios}
	\begin{split}
		\langle \hat{X} \rangle &= \langle \phi | \hat{X} |\phi\rangle \\
		\langle (\Delta \hat{X})^2 \rangle &= \langle \phi | (\hat{X}-\langle \hat{X} \rangle)^2|\phi\rangle.
	\end{split} 
\end{equation} 

One might then ask: if $\hat{X}$ is not an observable, does (\ref{valoresmedios}) still make sense? Of course we can still calculate these quantities (they are simply the matrix elements of an hermitean operator), but how could we interpret them? To answer this, we first note that, though $\hat{X}$ loses its status of an observable, it still has the physical interpretation of being an operator associated to the position of the system\footnote{An operator does not need to be an observable, not even hermitean(!), to have a physical interpretation. Take the creation and annihilation operators, $\hat{a}$ and $\hat{a}^{\dagger}$, as examples.}. That $\hat{X}$ is not an observable simply tells us that we cannot measure this position anymore (even to formulate the statement ``one cannot measure position because $\hat{X}$ is not an observable'' we already assume the relation between this operator and the spatial localization of the system.). Therefore, these mean values do tell us something about the way the system is distributed in space. 

Of course, we cannot interpret $\langle\hat{X}\rangle$ as the mean of the results obtained in a series of measurements, nor can we say that $\Delta x$ is the mean deviation of these measurements. But we can use a less statistical and more geometrical interpretation, associating $\langle \hat{X} \rangle$ to the point around which the system is localized, and $\langle (\Delta \hat{X})^2 \rangle$ to how much it is spread around this point.

The existence of a $\Delta x_{min}$ guarantees, therefore, that the physical system is never concentrated around a point with an arbitrary precision, in agreement with our interpretation of $\Delta x_{min}$ as a maximal resolution to the localization of the system.


\section{Minimal Length as a Regulator in QFT}
\label{regulator}

We now turn our attention to an extremely interesting property of a minimal length: its role as a natural regulator of Quantum Field Theory.

As already noted before, the notion of a particle perfectly localized in a point is not well defined in Quantum Mechanics, for the vector associated to such state does not belong to the space of physical states. In the non-relativistic theory this can be worked around by doing the calculations with the states $|\xi^{\Delta x}\rangle$, and taking the limit $\Delta x \rightarrow 0$ afterwards. We say that this is only a workaround because, though we recognize that the notion of precise localizability leads us to problems, we end up taking the limit that corresponds to such case, thus running from the responsibility of really solving the problem in its roots.

Of course, this can only be a temporary solution, and, indeed, this problematic notion of perfect localizability ends up giving rise to the much graver problem of the divergences that plague Quantum Field Theory. Indeed, it is now widely accepted that these divergences appear because, in QFT, the interactions are considered to occur in a single \emph{spacetime point}, or, in other words, because we evaluate the propagators as $G(x,x') = \langle x|\hat{G}|x' \rangle$\cite{Kempf:1998gk}.

Again, this difficulty can be bypassed by regularization methods similar to the procedure we did in the non-relativistic case, i.e. by introducing a restriction to the localizability of the interactions, doing the calculations and then taking the appropriate limit to return to the case of interactions occurring at a point. But, again, this mere workaround shows itself as unsatisfactory, because it does not work for the gravitational interaction --- to mention only one strong reason.

In this work we attack the very cause of these problems by introducing the notion of a nonzero minimal observable length, thus eliminating the notion of a perfectly localized event. In such case, the states $|\psi^{ml}_{\ \xi}\rangle$ maximally localized around a point $\xi$ are not eigenstates of the position operator, but are given by (\ref{DGSstate}), and the propagator must then be defined as $G(\xi,\xi ') = \langle \psi^{ml}_{\ \xi}|\hat{G}|\psi^{ml}_{\ \xi '} \rangle$. This introduces a non-local aspect to the field interactions, and we therefore expect the theory to be regular, i.e. divergence-free.

In A. Kempf\cite{Kempf:1996ss} it was shown that, for a scalar field theory with a $\phi^4$ self-interaction, this is, indeed, what happens: due to the presence of a minimal length, the Feynman diagrams do not diverge at any order of expansion!

In the following, we will also show that the Casimir Energy calculated from the formalism we have developed is finite. This is interesting because the infinities that appear in the calculations of the Casimir Effect are one of the most classic examples of the divergences that plague QFT, and it is a very strong result that the presence of a minimal length regularizes it. Moreover, due to high precision experimental measurements achieved, the Casimir effect may provide experimental constraints on the value of the minimal length\cite{panella,panella:2012,nouicer,Harbach:2005yu}.


\subsection{Casimir Effect}

Let us consider the problem of the one-dimensional Casimir Effect generated by the electromagnetic field in the presence of two parallel conducting ``plates'' separated by a distance $L$.

The Hamiltonian of the classical electromagnetic field is given by\cite{Plunien,Ryder}
\begin{equation}
	\label{hamiltonianaEB}
	H = \frac{1}{2} \int d^{3}x \left[ E^{2} + B^{2} \right] .
\end{equation}
Now, using the vector field $A^{\mu}$ and choosing the gauge condition $\vec{\nabla} \cdot \vec{A} = 0$ and $A^{0} = 0$ (transverse or Coulomb gauge) the Hamiltonian becomes
\begin{equation}
	\label{hamiltonianaA}
	H = \frac{1}{2} \int d^{3}x \left[ \left( \partial_{0} \vec{A} \right)^{2} +  \vec{A} \cdot \nabla^{2} \vec{A} \right] .
\end{equation}

In a model of electrodynamics in which the number of special dimensions is reduced from 3 to 1, the vector field $A^{\mu}$ is replaced with a scalar field $\phi$\cite{Boozer} and the Hamiltonian becomes, after an integration by parts,
\begin{equation}
	\label{hamiltoniana}
	H = \frac{1}{2} \int dx \left[ \left( \frac{\partial\phi}{c\partial t} \right)^{2} + \left( \frac{\partial \phi}{\partial x} \right)^{2} \right] .
\end{equation}

We will calculate the Casimir energy using the same approach of A. M. Frassino and O. Panella\cite{panella,panella:2012} and Kh. Nouicer\cite{nouicer}, that is, by performing a quantization procedure of the field. Because of the existence of the minimal length, instead of expanding the field operator in plane waves (which are not physical states anymore, as discussed above), we expand the field in the maximally localized states. From Eq. (\ref{DGSstate}) we have
\begin{equation}
	\label{DGSstate1}
	\psi^{ml}_{\xi}(k) = \sqrt{2\sqrt{ \frac{\beta} {\pi}}} e^{-i \xi k(p)} \cos\left( \hbar \sqrt{\pi \beta} k(p) \right).
\end{equation}
In this way, we obtain the field operator as
\begin{equation}
	\hat{\phi}(x,t) = \int \frac{dk}{\sqrt{2 \pi 2 \omega_{k}}} \cos \left( \hbar \sqrt{\pi \beta} k(p) \right) 
	\left[ \hat{a}_{k}e^{-i(kx -\omega_{k}t)} + \hat{a}^{+}_{k}e^{-i(kx -\omega_{k}t)} \right].
\end{equation}
	
The vacuum energy is given by
\begin{equation}
	E_{0} = \langle 0 | \hat{H} |0 \rangle ,
\end{equation}
with the Hamiltonian operator obtained from Eq.(\ref{hamiltoniana}).

After some algebra, we arrive at
\begin{equation}
	E_{0} = \frac{L}{4 \pi c} \int dk\ k\cos^{2} \left( \hbar\sqrt{\pi \beta }k \right) .
\end{equation}

The Casimir energy is the energy shift resulting from the presence of the two parallel conducting ``plates'', then
\begin{equation}
	E_{cas} = \langle 0 | \hat{H}(\Sigma) - \hat{H} |0 \rangle ,
\end{equation}
where $\hat{H}(\Sigma)$ is the Hamiltonian operator in the presence of the ``plates''.

The boundary conditions require that $k = \frac{n \pi}{L}$, for $n$ a positive integer. Now, from Eq.(\ref{relationkp}) we see that the wave-number $k$ has a maximum value given by\footnote{The functional relation (\ref{relationkp}) between the wave-vector $k$ and the momentum $p$ is one that has been used by U. Harbach and S. Hossenfelder in order to numerically calculate the Casimir energy in the presence of a minimal length\cite{Harbach:2005yu}.}
\begin{equation}
	k_{max} = \frac{1}{2 \hbar} \sqrt{\frac{\pi}{\beta}},
\end{equation}
so there is a corresponding maximum value $n_{max}$, given by the greatest integer smaller than $\frac{L}{2\hbar \sqrt{\pi \beta}}$. Therefore,
\begin{equation}
	\label{casimir_energy}
	E_{cas} = \frac{\pi}{2Lc}\sum_{n=0}^{n_{max}} \left[ n \cos^{2} \left(\frac{\hbar\sqrt{\pi\beta}\pi}{L} n \right) \right] - 
	\frac{\pi}{2Lc}\int_{0}^{\nu_{max}} \nu \cos^{2} \left(\frac{\hbar\sqrt{\pi\beta}\pi}{L} \nu \right) d\nu ,
\end{equation}
where $\nu_{max} = \frac{L}{2\hbar \sqrt{\pi \beta}}$. The finiteness of the result becomes clear, since the integral converges (!) and the sum is over a finite number of terms. This shows that a minimal length indeed acts as a regulator of QFT, as discussed above. Note that in the case of free space we have taken the plates to infinity, so the discrete variable $n$ becomes a continuous one, $\nu$.

We can calculate the energy explicitly using the Euler-Maclaurin formula, according to which
\begin{equation}
	\label{Euler-Maclaurin}
	\sum_{k=0}^n G(k)=\int_0^n G(x)dx +\frac{1}{2} \left[G(n) + G(0) \right] +\sum_{k=1}^{N}\frac{B_{2k}}{(2k)!} \left[ G^{(2k-1)}(n)-G^{(2k-1)}(0) \right] +R_N ,
\end{equation}
where $B_{2k}$ are the Bernoulli numbers, $N$ is an arbitrary integer and $R_N$ is the error of the approximation for a given $N$. However, it is difficult to analyse the result obtained using this formula, since we will end up with a functional dependence on $n_{max}$, and we do not have a closed expression for this variable as a function of $L$, $\beta$ and other parameters of the problem. 

To work around this difficulty and perform the calculations, we will use a small trick. Note that, since the integral in (\ref{casimir_energy}) converges, for every $\epsilon>0$ we choose there is a $\delta\in(0,\nu_{max})$ such that $\int_0^{\delta} \nu \cos^{2} \left( \frac{\hbar\sqrt{\pi\beta}\pi}{L} \nu  \right)d\nu$ differs from the desired integral by less than $\epsilon$. Thus, given an $\epsilon>0$, let us define $G(\nu)$ as a function that equals $ \nu \cos^{2} \left( \frac{\hbar\sqrt{\pi\beta}\pi}{L} \nu \right)$ for all $\nu \leq  max\{\delta,n_{max}\}$, goes smoothly to zero for $\nu > max\{\delta,n_{max}\}$, and equals zero for every $\nu \geq \frac{L}{2 \hbar \sqrt{\pi \beta}}$. We then have\footnote{We can extend the sum to infinity because, by definition of $n_{max}$, $n>n_{max} \Rightarrow n>\frac{L}{2\hbar \sqrt{\pi \beta}} \Rightarrow G(n)=0$.}
\begin{equation}
	\sum_{n=0}^{n_{max}}  n\cos^{2} \left( \frac{\hbar\sqrt{\pi\beta}\pi}{L} n \right) = \sum_{n=0}^{n_{max}} G(n) = \sum_{n=0}^{\infty} G(n)
\end{equation}
and
\begin{equation}
	\int_0^{\frac{L}{2 \hbar \sqrt{\pi \beta}}} \nu \cos^{2} \left( \frac{\hbar\sqrt{\pi\beta}\pi}{L} \nu \right) d\nu - \int_0^{\infty}G(\nu)<\epsilon,
\end{equation}
and we can therefore write
\begin{equation}
	E_{cas} = \frac{\pi}{2Lc}\sum_{n=0}^{\infty} G(n) -	\frac{\pi}{2Lc} \int_{0}^{\infty} G(\nu) d\nu .
\end{equation}
up to an error $\epsilon$ that we control. This extension of the domain of the sum/integral to infinite was our motivation to introduce the new function $G(\nu)$. Since $G(\nu)=0\ \forall\ \nu \geq \frac{l}{2 \hbar \sqrt{ \pi \beta}}$, and because $G(\nu)=  \nu \cos^{2} \left( \hbar \sqrt{\pi \beta} \nu \right)$ in a neighborhood of $\nu=0$, Eq. (\ref{Euler-Maclaurin}) leads us to
\begin{equation}
	E_{cas}= -\frac{\pi}{2Lc}\sum_{k=1}^{\infty}\frac{B_{2k}}{(2k)!}\frac{d^{2k-1}}{d\nu^{2k-1}} \nu \cos^{2} \left( \frac{\hbar\sqrt{\pi\beta}\pi}{L} \nu \right)\bigg|_{\nu=0} ,
\end{equation}
and we thus arrive at a powerful expression that allows us to compute the Casimir energy to any order we want.

Evaluating the derivatives using $B_{2} = \frac{1}{6}$ and $B_{4} = -\frac{1}{30}$ we obtain
\begin{equation}
	\label{e2}
	E_{cas}=-\frac{\hbar\pi c}{24L} - \frac{\hbar^3\pi^4 c}{240 L^3}\beta + \mathcal{O}(\beta^2).
\end{equation}

The Casimir force between the "plates", given by $F_{cas} = - \frac{\partial E_{cas}}{\partial L}$, is, therefore,
\begin{equation}
	\label{e2}
	F_{cas}=-\frac{\hbar\pi c}{24L^2} - \frac{\hbar^3\pi^4 c}{80 L^4}\beta + \mathcal{O}(\beta^2).
\end{equation}
Note  that, to $ \mathcal{O}(\beta^0)$, we recover the well-known L\"uscher result for the (1+1)-dimensional Casimir Force \cite{Luscher:1980,Luscher:1981}. We also note that second term in Eq. (\ref{e2}) is negative, so the minimal length does not change the sign of the Casimir force. Similar result has been obtained by A. M. Frassino and O. Panella\cite{panella:2012}.


\section{Conclusions and Outlooks}
\label{conclusions}

The introduction of a minimum length has been carried out in various ways, among them by modifying the commutation relations, by modifying the dispersion relation and the non-commutativity of components of spacetime. In this paper we have considered a modification of the commutation relation between the position and momentum operators in the form
\begin{equation}
	\label{algebraproposta2}
	[\hat{X},\hat{P}]=i\hbar e^{\beta \hat{P}^2}\ ,
\end{equation}
with the intention of taking into account all orders of minimum length parameter $ \beta $. Initially we showed that in fact the modifided commutation relation (\ref{algebraproposta2}) leads to the introduction of a minimal length. As was shown by Detournay, Gabriel and Spindel\cite{Detournay:2002fq}, in the case of the commutation relation (\ref{algebraproposta2}) the prescription of Kempf, Mangano and Mann\cite{Kempf:1994su} is not appropriate for calculation of the maximally localized states. Therefore, we used the DGS prescription to calculate the maximally localized states and the minimal length.

We have stressed that a minimal length acts as a natural regulator of Quantum Field Theories, so that if we manage to extend the formalism here developed to QFT we end up with a divergence-free theory. Then  we not only showed that the minimal length regularizes the Casimir Energy, we have also calculated it. We have found that the correction of order $\beta$ to the Casimir force is attractive,when the Casimir energy is computed by expanding the field in the maximally localized states that we have calculated. Our result is in agreement with that of A. M. Frassino and O. Panella\cite{panella:2012}.

There is also another consequence of the existence of a minimal length that we have not discussed in this work, but which is also worth mentioning. Let $\ell_{min}$ be a minimal length, and suppose we have a certain object of length $\ell_{min}$, as measured in the rest frame $S$ of the object. Then, if we accept the results of Special Relativity (in particular, Lorentz contraction), a frame $S'$ moving with respect to $S$ would measure the length of the object to be $\ell<\ell_{min}$, which is absurd. This means that, if we accept the existence of a minimal length, the postulates of the theory of Relativity must be modified to avoid this contradiction.

In \cite{DoublyRelativity,Doubly2,Doubly3,Doubly4,Magueijo:2002am} it was shown that this can be done by including a postulate according to which the minimal length $\ell_{min}$ is also a universal constant of nature, together with $c$. We then have a theory of Relativity with two universal constants --- hence the name Doubly Relativity. The resulting expression for the ``generalized Lorentz contraction'' is such that the length $\ell$ of a moving object is always greater than or equal to $\ell_{min}$, regardless of its speed relative to the observer, and the above mentioned inconsistency is, therefore, solved.

In this work, we have restricted ourselves to the simple one-dimensional case. With more than one spatial dimension, the more general form of the commutation relation which depends only on the momentum up to quadratic orders and respects rotational symmetry can be written as\cite{chang}
\begin{equation}
	\label{rc3dg}
	[\hat{X}_i,\hat{P}_j] = i\hbar \left[ A \left(\hat{\textbf{P}}^2 \right) \delta_{ij} + B \left(\hat{\textbf{P}}^2 \right)\hat{P}_{i}\hat{P}_{j} \right],
\end{equation}
where $\hat{\textbf{P}}^2 = \sum_{i} \hat{P}^{2}_{i}$ .

Various choices for the functions $A \left(\hat{\textbf{P}}^2 \right)$ and $B \left(\hat{\textbf{P}}^2 \right)$ have been considered in the literature. A. Kempf\cite{Kempf:1997,Kempf2:1997} considered the functions $A \left(\hat{\textbf{P}}^2 \right)$ and $B \left(\hat{\textbf{P}}^2 \right)$ so that the commutation relation is given by
\begin{equation}
	\label{rc3kempf}
	[\hat{X}_i,\hat{P}_j] = i\hbar \left[  \left(1 + \beta\hat{\textbf{P}}^2 \right) \delta_{ij} + \beta^{\prime}\hat{P}_{i}\hat{P}_{j} \right],
\end{equation}
where $\beta^{\prime}$ is other parameter related to the minimal length.
If the components of the momentum operator are assumed to commute with each other,
\begin{equation}
	\label{rcM}
	[\hat{P}_i,\hat{P}_j] = 0,
\end{equation}
then the commutation relations among the components of the position operator are determined by the Jacobi identy as\footnote{From now on there is a summation over dummy indices.}
\begin{equation}
	\label{rc2kempf}
	[\hat{X}_i,\hat{X}_j] = -i\hbar \left[ 2\beta - \beta^{\prime} + \left(2\beta + \beta^{\prime} \right)\beta\hat{\textbf{P}}^2 \right] \epsilon_{ijk}\hat{L}_{k},
\end{equation}
where
\begin{equation}
	\label{rc3kempf}
	\hat{L}_{i} = \frac{1}{\left(1+ \beta\hat{\textbf{P}}^2 \right)} \epsilon_{ijk}\hat{X}_{j}\hat{P}_{k},
\end{equation}
are the components of the orbital angular momentum operator, satisfying the usual commutation relations $[\hat{L}_{i},\hat{X}_{j}] = i \hbar \epsilon_{ijk}\hat{X}_{k}$ and $[\hat{L}_{i},\hat{P}_{j}] = i \hbar \epsilon_{ijk}\hat{P}_{k}$. This algebra gives rise to (isotropic) nonzero minimal uncertainties in the position coordinates $\Delta X^{min}_{i} = \hbar \sqrt{3 \beta + \beta^{\prime}}$, as it was expected. However this algebra is not Lorentz covariant. C. Quesne and V. M. Tkachuk\cite{Quesne1,Quesne2} have proposed a generalization of the Kempf algebra to a Lorentz-covariant algebra,
\begin{equation}
	\label{quesne1}
	[\hat{X}^{\mu},\hat{P}^{\nu}] = -i\hbar \left[  \left(1 - \beta\hat{P}_{\rho}\hat{P}^{\rho} \right) g^{\mu \nu} - \beta^{\prime}\hat{P}^{\mu}\hat{P}^{\nu} \right],
\end{equation}
\begin{equation}
	\label{quesne2}
	[\hat{X}^{\mu},\hat{X}^{\nu}] = i\hbar \frac{\left[ 2\beta - \beta^{\prime} + \left(2\beta + \beta^{\prime} \right) \beta\hat{P}_{\rho}\hat{P}^{\rho} \right]} { \left(1 - \beta\hat{P}_{\rho}\hat{P}^{\rho} \right)}
 \left(\hat{P}^{\mu}\hat{X}^{\nu} - \hat{P}^{\nu}\hat{X}^{\nu} \right),
\end{equation}
\begin{equation}
	\label{quesne3}
	[\hat{P}^{\mu},\hat{P}^{\nu}] = 0,
\end{equation}
 which includes Snyder algebra\footnote{H. S. Snyder\cite{Snyder} assumed that the space-time is no continuous, this leading to a Lorentz-covariant quantized space-time, where the components of the position no longer commute with each other. Thus a minimal length is implemented. } as a special case where $\beta = 0$. It is worth noting that the Kempf algebra can not be obtained by taking the nonrelativistic limit of the algebra proposed by Quesne and Tkachuk. From all this, we can see that a generalization to the tridimensional case which considers all orders of the minimum length parameter must be developed carefully. Moreover, a quick glance at Eq.(\ref{rc2kempf}) or Eq.(\ref{quesne2}) shows that in the special case $\beta^{\prime} = 2\beta$ the components of the position operator only commute to first order of the minimum length parameter. Therefore, it is expected that an approach that includes higher orders in $\beta$ and which recovers the one of Kempf or the one of Quesne and Tkachuck to first order incorporates a non-commutative spacetime\cite{Doplicher:1994tu,Doplicher:1994zv}.

Finally, it would be interesting to extend this formalism from the non-relativistic case to QFT\footnote{F. Girelli and Etera R. Livine have constructed a scalar field theory on Snyder space-time\cite{Girelli1}.}. This is not simple, because we must take into account not only the change in the commutation relations, but also in the invariance group of the theory, which is not the Poincar\'e group anymore (because of the comments made a few paragraphs above), but a more complicated extension of it\cite{Girelli2}. It would also be interesting to perform the calculations of the Casimir Force for the three-dimensional case, and compare the results obtained with the experimental data, which would give us an upper bound for the minimal length. 


\section*{Acknowledgements}

We would like to thank CAPES (Brazil) for financial support.



\end{document}